\documentclass[reprint,aps,prl,showpacs,twocolumn]{revtex4}
\usepackage{epsf,graphicx}

\usepackage{listings}
\usepackage{amssymb}
\usepackage{eso-pic}
\usepackage{booktabs}

\begin{document}
\vspace*{1cm}
\renewcommand{\thefootnote}{}
\title{Cystine plug and other novel mechanisms of large mechanical stability in dimeric proteins}
\author{Mateusz Sikora}
\affiliation{Institute of Physics, Polish Academy of Sciences\\
Al. Lotnik\'ow 32/46, 02-668 Warsaw, Poland}
\author{Marek Cieplak}
\affiliation{Institute of Physics, Polish Academy of Sciences\\
Al. Lotnik\'ow 32/46, 02-668 Warsaw, Poland}
\email{mc@ifpan.edu.pl}

\begin{abstract}
We identify three dimeric proteins whose mechanostability is anisotropic and
should exceed 1 nN along some directions.
They come with distinct mechanical clamps: shear-based, involving
a cystine slipknot, and due to dragging of a cystine plug through a cystine ring.
The latter two mechanisms are topological in nature and the cystine plug mechanism
has not yet been discussed but it turns out to provide the largest
resistance to stretching. Its possible applications in elastomers are discussed.
\end{abstract}
\maketitle

\vspace*{1cm}

Cell-cell adhesion, protein translocation, muscle extension, activation of
mechanosensory pathways, switching on of catalytic functions of proteins,
and other biological processes involve protein unfolding as a
result of action of a force \cite{Crampton,encyclopedia,vinculin,titin,Vogel}.
Unfolding forces, $F_{max}$, usually range from 10  to 300 pN \cite{Oberhauser,PLOS}.
$F_{max}$ may be defined as tension corresponding to the largest force peak
identified during stretching at constant speed on the way to the total
unraveling of the tertiary structure.
Several proteins, however, have been found to have larger 
$F_{max}$: two types of scaffoldins have $F_{max}$ of 425 and 480 pN \cite{Valbuena},
a certain way of stretching of the green fluorescent protein yields 548 pN \cite{Dietz},
and protein molecules in the spider capture-silk
have $F_{max}$ as high as 800-900 pN \cite{Hansma} (but 176 $\pm$ 73 pN
in the spider dragline \cite{Oroudjev}).
Are there proteins with $F_{max}$ exceeding 1000 pN and can such stability be harnessed?

Here, we identify three examples of proteins that, based on simulations,
should be very robust mechanically. They are all dimeric so their resistance to stretching
is very anisotropic: certain 
ways to implement pulling are very hard and other are easy.
Interestingly, their responses to stretching are governed by three different
mechanisms, or mechanical clamps, one of which has not yet been
identified before this work. The first of these proteins is ATU1913 from
{\it Agrobacterium tumefaciens} strain C58 \cite{Joachimiak} with an unknown function.
Its Protein Data Bank (PDB) \cite{Berman2000} structure code is 2B1Y.
The second is a neurotrophic growth factor artemin with the PDB code 2GH0
\cite{2GH0} and high thermal stability \cite{artemin}.
The third is the human transforming growth factor-$\beta$2 \cite{Schlunegger}
with the PDB code 1TFG. The transforming growth factor
binds to various receptors easily and is involved
in wound healing, bone formation, and modulation of immune functions.
It is in this protein that we find a novel type of the mechanical clamp.

\begin{figure}[h!]
\begin{center}
\includegraphics[width=0.5\textwidth]{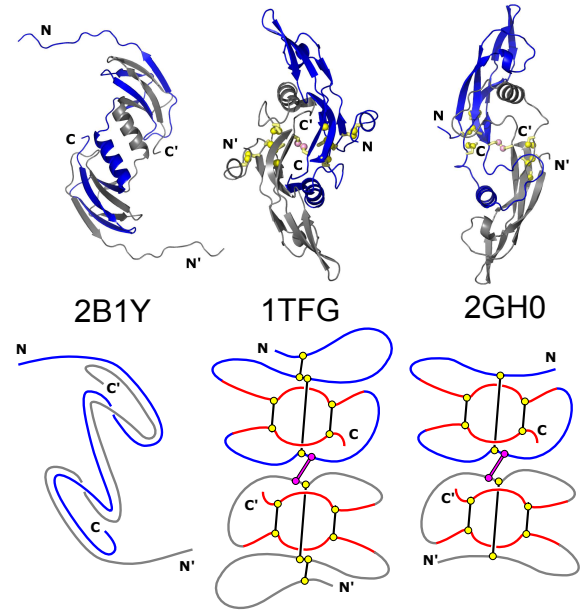}
\caption{{\small Top panels: 
structural representations of the proteins studied here.
Bottom panels: simplified versions of the structures -- they
illustrate the nature of connectivities.
The brighter (yellow) circles
correspond to the atoms of sulfur belonging to the cystine rings.
The darker (magenta) circles show these atoms in cystines that
link the monomers. The termini and secondary structures 
are indicated. The unprimed symbols refer to one monomer and the primed symbols
to the other.
In the lower panels,
symbols N,C and N', C' point to residues which are sequentially
closest to the indicated termini.
The intra-monomer cystines are represented by thick black lines, whereas
the inter-monomer bridges are in brighter lines. 
The red lines highlight vicinity of a cystine ring.
}}
\label{dimers}
\end{center}
\end{figure}

\begin{figure}[htb]
\begin{center}
\includegraphics[width=0.5\textwidth]{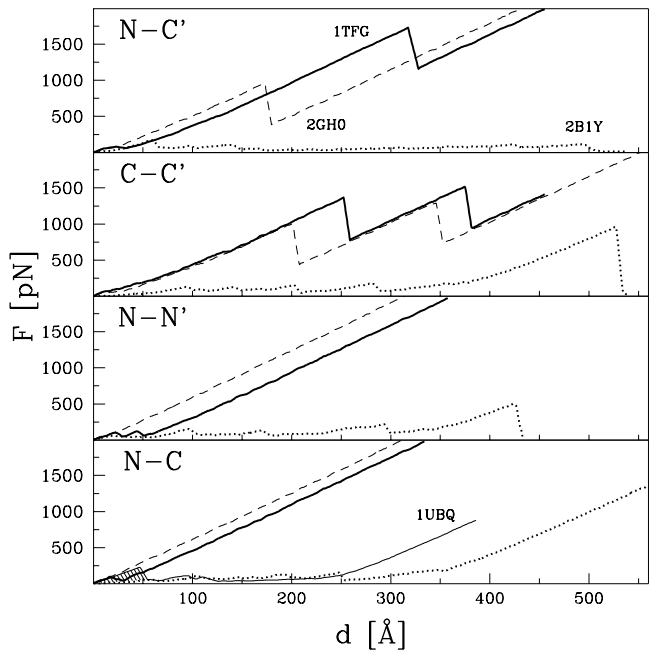}
\caption{ {\small The $F-d$ curves for the proteins studied in this paper.
The ways of pulling are indicated in the upper left corner of each panel.
The line type for a given protein is the same throughout.
The thin solid line in the bottom panel is for ubiquitin (a monomer).
The shaded area under this line gives the energy needed
to extend the protein just past the first peak.
}
} \label{force}
\end{center}
\end{figure}

A schematic representation of the native structures of these proteins is shown
in figure \ref{dimers}. The termini in one monomer are denoted by N and C. In
another -- by N' and C'. Mechanostability in an experiment involving single molecule
manipulation \cite{Nagy} can be measured by anchoring one terminus, pulling by another
at constant speed $v_p$ (here: $v_p\sim 5\;10^5$ nm/s -- the speed for which
most of our previous studies were made \cite{PLOS}; both termini
are attached to elastic elements), and by determining the height of the
largest force peak associated with a conformational transformation.
The plots of the tension force, $F$, vs. moving end displacement, $d$,
are shown in figure \ref{force}. The values of $F_{max}$ are
summarized in Table I. After the generation of the force peaks is completed,
the tension grows monotonically since only the covalent bonds (along the
backbone and in the disulfide bonds) resist the manipulation.

\begin{table}[h!]
\caption{{\small Values of $F_{max}$ of the three dimeric proteins studied here in pN
and for different pulling schemes.
$F_M$ denotes $F_{max}$ in the monomeric case -- when only
one chain of the dimer is considered in the N-C scheme.
The remaining columns are for the the dimeric situation. The subscripts of $F$
indicate the mode of pulling. Two values of $F_{max}$ are listed for each scheme.
The first one is for $v_p \sim 5\;10^5$ nm/s and the second is a logarithmic
extrapolation to $v_p$=500 nm/s. The extrapolation is based on calculating $F_{max}$
for five values of $v_p$ and using $F_{max} = q + p\;ln(v_p/v_0)$
with $v_0=10^8$ nm/s.}\\  \\
}
\label{tabela1}

\begin{tabular}{@{}l|ll|ll|ll|ll|ll@{}}
\multicolumn{1}{c}{PDBid} & \multicolumn{2}{|c}{$F_{M}$}& \multicolumn{2}{|c}{$F_{N-C'}$} & \multicolumn{2}{|c}{$F_{N-N'}$} & \multicolumn{2}{|c}{$F_{C-C'}$}& \multicolumn{2}{|c}{$F_{N-C}$} \\
\cmidrule[0.8pt](r){1-11}
2B1Y                &  45  & 20   & 190     &145                   &  530 &420                      & $\;$990 &870                  &  165 &70  \\
2GH0                &  650 & 430   & 950     &810                   & $-$  &$-$                      &1320     &1150                 &  $-$&$-$  \\
1TFG                &  650 & 630   & 1560    &1280                  & 120  &90                       &1540     &1250                 &  120&75  \\

\end{tabular}
\end{table}



\begin{figure}[htb]
\begin{center}
\includegraphics[width=0.5\textwidth]{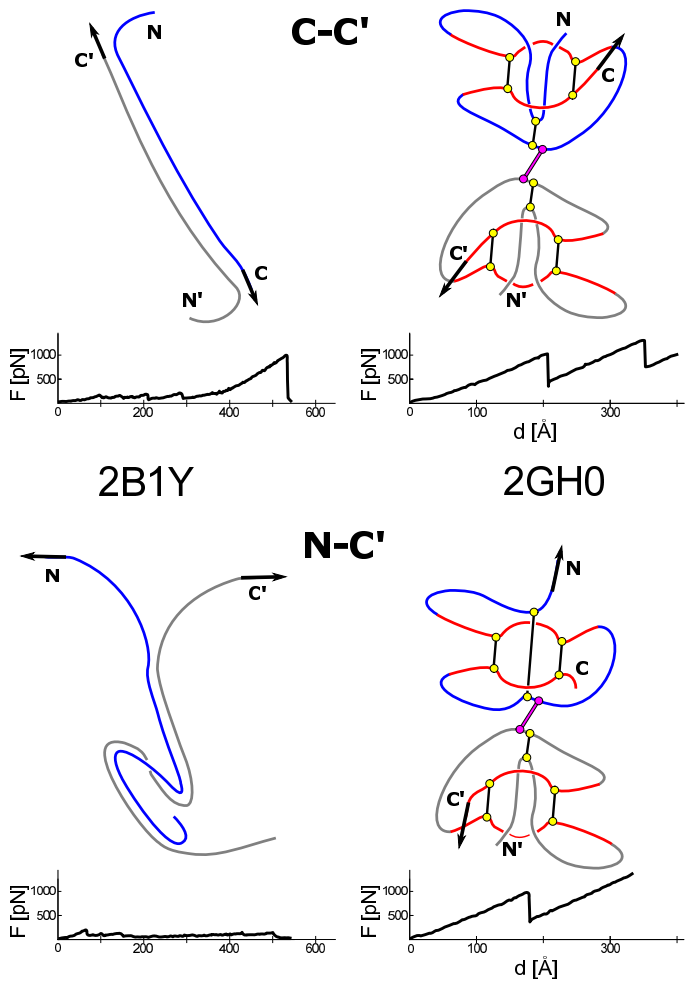}
\caption{{\small Mechanisms involved in stretching of protein 2B1Y (the left panels)
and 2GH0 (the right panels). The upper (lower) panels are for the C-C' (N-C') pulling.
The drawings represent conformations at $F_{max}$.
The corresponding $F-d$ plots are shown underneath.
}} \label{mechanism}
\end{center}
\end{figure}

Pulling of single chains by the termini is distinguished from pulling by other choices
of force attachment since it leads to unravelling of all parts of the tertiary
structure and thus comes (usually) with the the largest $F_{max}$. In addition,
it is often implemented experimentally through attachment to flanking
reference proteins. For symmetric dimers
there are four choices of pairs of the termini: 
N-C', C-C', N-N', and N-C.
The first three of these drive towards separation of the dimers (see figure \ref{mechanism}).
The monomers in 2B1Y are linked through contact interactions provided by hydrophobicity and
hydrogen bonds between two sets of two $\beta$-sheets and one helix (figure \ref{dimers}).
The force peaks arising during the
process result from shear in contacts within the monomers and between the monomers. The
shear-based mechanical clamps have been first identified 
\cite{Schulten} in the context of titin and
ubiquitin for which the measured value of $F_{max}$ are close to 200 pN \cite{titin}.
Figure \ref{force} demonstrates that the C-C' pulling of 2B1Y may yield $F_{max}$
close to 1000 pN.
This value of the force exceeds $F_{max}$ of about 770 pN predicted by us for the dimeric
(3D-domain swapped) cystatin C \cite{domains}
-- for the N-N' pulling at the same speed -- in which the shear mechanical clamp is also
operational. $F_{max}$ of 2B1Y gets halved for the N-N' pulling and becomes still smaller
for N-C and N-C' as then one observes mostly unzipping of the strands as illustrated
in figure \ref{mechanism}.
The simulations have been performed within a coarse-grained molecular dynamics model
constructed empirically based on the knowledge of the native structure \cite{JPCM,thermtit}.
The model uses parameter $\epsilon$ which denotes the
depth of the potential representing native contacts between the C$^{\alpha}$ atoms.
It also provides the amplitude for soft repulsion in non-native contacts.
We have used the callibration of $\epsilon$/{\AA} of around 110 pN  derived by
making comparisons to experimental data on stretching \cite{PLOS}.
The list of the native contacts is derived through studies of the geometric overlaps
between spheres representing amino-acidic atoms and related to their van der Waals volumes.
The value of $F_{max}$ depends  on the temperature, $T$.
Our simulations are performed at $k_BT/\epsilon$=0.35 which should represent
the room temperature behavior ($k_B$ is the Boltzmann constant).
Effects of the solvent are included in an implicit way.

The structures of monomers in 2GH0 and 1TFG, in addition to $\alpha$-helices
and $\beta$-sheets (figure \ref{dimers}), include a distinct and tight
motif known as the cystine knot \cite{Wlodawer,Bradshaw,Sun,Iyer}.
It involves three disulfide bonds which connect three pairs of cysteins, forming three
cystins. Two of these close a covalently linked ring made of eight residues. The
third points through the ring so that one segment of the backbone is above
and another below the ring. This motif is found in many growth factors.
The 2GH0 and 1TFG dimers are formed by linking to monomers near the connections
through the rings by still another disulfide bond. Such covalently bound dimers
(unlike the case of 2B1Y) cannot separate
unless one applies a force of, probably, some 4 nN or more. And yet
conformational changes may take place. They correspond to well defined force
peaks on the plot of  $F$, vs.  $d$ shown in figure \ref{force}.
The most costly of them defines $F_{max}$.

The force peak in 2GH0 has been interpreted recently as being due to formation
of a cystine slipknot mechanical clamp \cite{PLOS,NAR}. This clamp involves dragging
of the backbone (acting as a knot-loop) through the tight cystine ring until no further
relative motion is possible. The argument has been based on
simulations within the coarse-grained model and the findings
have been also confirmed by all-atom simulations as well \cite{Peplowski}. However,
these calculation have considered just one monomer instead of taking the dimeric
nature of the protein into account. Here, we demonstrate that the slipknot mechanism
is valid, but the behavior is richer. When pulling in the N-C' way, $F_{max}$ is
1.5 times larger than the monomeric value, and twice as large -- about 1300 pN --
when pulling in the C-C' way. For the N-N' and N-C stretching, $F$ just grows monotonically
with $d$ which signifies overall distortion of the system without any well defined
conformational change.

Finally, our simulations suggest that the third of the proteins studied here, 1TFG,
behaves similar to 2GH0. 
The spring constant of the initial response to the relative termini
motion in the C-C' pulling is 0.064 N/m for 1TFG and 0.034 N/m for 2GH0. 
Both values are larger than that of the spider dragline (0.015 N/m)
\cite{Oroudjev} and a similar relationship holds for other ways of
pulling. There is, however, one important difference:
the knot-loop dragged through the ring by a driving cysteine belongs
to a bulky cystine plug. The plug is a loop that is closed by a disulfide bond
so that another effective ring is formed. The plug comprises 10 residues and
is just bigger than the cystine ring. It is near the N-terminus
(and N') of the protein. The resulting values of $F_{max}$ are predicted to be
about 1500 pN both for the N-C' and C-C'
manipulation and an order of magnitude less for the other two 
choices of the termini.
If confirmed experimentally, this would be the most potent mechanical clamp
in proteins. It is not clear how many other proteins are structured this way.

We now discuss some details of the mechanisms leading to the $F-d$ curves
shown in figure \ref{force}.
The two backbones in 2B1Y are intertwined (see figure \ref{dimers}) at the
nearly full length of each.
After rupturing secondary structures, stretching results
in an antiparallel positioning of the strands from the two monomers
(figure \ref{mechanism}). In the N-N' and C-C' modes, further pulling generates
large shear since many contacts get stretched.
The halved value of $F_{max}$ in the N-N' pulling  is a result 
of a less compact structure with fewer contacts near the C and C' termini.
The N-C' and N-C stretchings result in unzipping which generates weak forces.

The stretching mechanisms in 2GH0 are illustrated in figure \ref{mechanism}.
There are two major force peaks when 2GH0 is pulled in the C-C' way because
two slipknots form, but not simultaneously. For the N-C' pulling, only one
slipknot forms -- in the cystine ring belonging to the N'-C' monomer.
There is no dragging of the backbone through the other cystine ring because
the N terminus gets aligned with the cysteine providing the link to the
other (lower in the figure) monomer and the pulling force is
directly transmitted, through the ring that is pierced by the cystine,
to the second ring. The slipknot is formed there because C' is at an angle
relative to the direction of alignment.
There are only minor force peaks arise in the N-N' and N-C pullings (figure \ref{force}).
They are generated during the process of alignment.

\begin{figure}[htb]
\begin{center}
\includegraphics[width=0.5\textwidth]{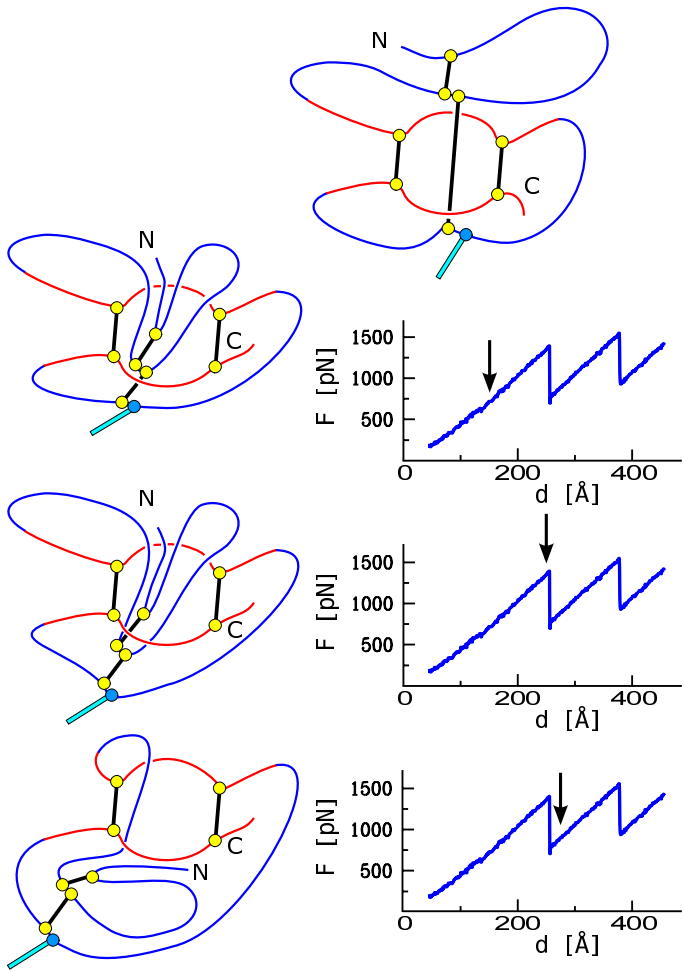}
\caption{{\small Emergence of the cystine plug mechanical clamp in 1TFG during
the C-C' stretching.  One monomer is shown fully 
and the other is indicated by its pulling fragment (in green).
The panel on top right shows the native conformation. The panels on the left - subsequent
conformations. The corresponding $F-d$ curves are shown in the right hand panels.
The arrow indicates the stage of pulling. There are two force peaks because a
similar squeezing through occurs at the other cystine ring later on.
}}
\label{double}
\end{center}
\end{figure}

Figure \ref{double} explains the C-C' puling in 1TFG.
It involves dragging of the cystine plug together with the short segment
connected to the N-terminus through the cystine ring. There are two force
peaks because the process is repeated at both rings. Each peak is
large (see Table I) because  both the ring and the plug
undergo substantial structural adjustments.

Driving the plug through the ring is rapid at takes place at a
large force. It results in discharging the plug at a great velocity.
We observe that the distance between the C and C' termini gets
tripled within a short time ($\sim$1 ns).
It should be possible to harness this mechanism
in biocompatible nanomachines designed to absorb energy,
or to trigger some action at a large strain, or locking  a system in a long-lived
metastable state.
Such nanomachines could be involved in biological shock absorbers or
in prostheses of limbs and tendons. The cystine plug proteins might also provide
"hooks" when making proteinic fabrics, wound dressings or sutures.
The action of
such a device could be either of a single use
or repetitive. 
For each of the three proteins studied here
the stretching process is reversible until the maximum of the first force peak
is reached. After crossing the peak, the behavior appears to be  irreversible:
either the monomers stay apart (the case of 2B1Y) or the plug does not
thread back despite waiting for more than a $\mu$s (1TFG). It is
expected, however, that reversibility is attained at sufficiently long times.

Proteins with high mechanical stability, when unfolding,
absorb large amounts of energy on a short path.
For ubiquitin, the work needed to just pass $F_{max}$
(the shaded area in the bottom panel of figure \ref{force}) is about 1 fJ.
For 1TFG, the cystine plug mechanism
makes this work much larger. Crossing just past the first peak
requires 160  fJ and the second -- 300 fJ.
For a mole of 1TFG, full unfolding would require 18 MJ of energy.
Such parameters are considerably higher than for proteins considered
in elastomeric applications \cite{Urry,Hongbin,Vassalli}.
Building elastomers requires connecting the units, in this case dimers of
1TFG or 2GH0, into chains. This can be accomplished
by extending the (buried) C-termini by peptide linkers and connecting
them either to N or C on the next dimer (the latter may need making
a mutation that allows for a formation of a linking cystine.
We find that subsequent plugs
in the chain are released in a serial fashion so there is
no compounding of $F_{max}$ but there is one in absorbed energy.

The magnitudes of $F_{max}$ predicted for the three dimers
are large yet still below the
tension needed to break covalent bonds \cite{Grandbois}.
Large breakage forces have been observed for rupture of heterogeneous
proteinic systems: the titin-telethonin complex at the Z-disk portion
of a skeletal muscle is bound by a force of order 707 pN \cite{Bertz}.
An experimental verification of our findings is a necessary prerequisite
for considering applications and the $v_p$-dependence of $F_{max}$
has to be taken into account. The values of $v_p$ in AFM experiments
vary between 300 and 12 000 nm/s \cite{PLOS}. We have performed calculations
for six speeds between $5\; 10^4$ and $10^7$ nm/s and found the results to be
compatible with a logarithmic $v_p$-dependence \cite{PLOS} in which the coefficient
of proportionality depends on the protein and on the pulling scheme
(see the Supplementary Information).
The extrapolated values of $F_{max}$ to $v_p$ of 500 nm/s, at the
lower end of possible experimental speeds, are listed in Table I.
Applications also require designing a
bio-compatible strategy of assembly of these proteins into longer chains.
The cystine plug mechanism could be an element of a sacrificial network.
Proteins 2GH0 and 1TFG have been singled out by making a survey of about
100 growth factors identified before \cite{PLOS} and augmented by several
similar structures. It would be worthwhile to explore other systems with
the cystine knots.

\vspace*{0.5cm}
\begin{acknowledgments}

We appreciate discussions with M. Carrion-Vazquez, D. Elbaum,
M. Vassalli, T. W{\l}odarski, and A. Wlodawer.
The computer resources were financed by the European Regional
Development Fund under the Operational Programme Innovative Economy
NanoFun POIG.02.02.00-00-025/09. This research has been supported by
the Polish National Science Centre Grants No. 2011/01/B/ST3/02190 (MC)
and 2011/01/N/ST3/02475 (MS).
\end{acknowledgments}

\end{document}